\documentclass[aip,reprint]{revtex4-2}

\usepackage{graphicx}
\usepackage{epsfig}
\usepackage{amsmath,amssymb}
\usepackage{bbm}
\usepackage{dcolumn}
\usepackage{bm}
\usepackage[title]{appendix}
\numberwithin{equation}{section}

\begin{document}

\preprint{APS/123-QED}

\title{Two-molecule theory of polyethylene liquids}
\author{Huimin Li}
\affiliation{Department of Chemical Engineering, Colorado School of Mines, Golden, CO 80401}
\author{James P. Donley}
\email[Corresponding author: ]{jpdonley7@icloud.com}
\affiliation{Material Science Institute, University of Oregon, Eugene, OR 97403}
\author{David T. Wu}
\email[Corresponding author: ]{dwu@mines.edu}
\affiliation{Department of Chemical Engineering, Colorado School of Mines, Golden, CO 80401}
\affiliation{Department of Chemistry, Colorado School of Mines, Golden, CO 80401}
\author{John G. Curro}
\affiliation{Department of Materials and Metallurgical Engineering, New Mexico Institute of Mining and Technology, Socorro, NM 87801}
\author{Caleb A. Tormey}
\affiliation{Department of Chemistry, The College of Idaho, Caldwell, ID 83605}

\date{\today}

\begin{abstract}
Two-molecule theory refers to a class of microscopic, self-consistent field theories for the radial distribution function in classical molecular liquids. The version examined here can be considered as one of the very few formally derived closures to the reference interaction site model (RISM) equation. The theory is applied to polyethylene liquids, computing their equilibrium structural and thermodynamic properties at melt densities. The equation for the radial distribution function, which is represented as an average over the accessible states of two molecules in an external field that mimics the effects of the other molecules in the liquid, is computed by Monte Carlo simulation along with the intramolecular structure function. An improved direct sampling algorithm is utilized to speed the equilibration. Polyethylene chains of 24 and 66 united atom $\rm CH_2$ units are studied. Results are compared with full, many-chain molecular dynamics (MD) simulations and self-consistent polymer-RISM (PRISM) theory with the atomic Percus-Yevick (PY) closure under the same conditions. It is shown that the two-molecule theory produces results that are close to those of MD, and is thus able to overcome defects of PRISM-PY theory and predict more accurate liquid structure at both short and long range.  Predictions for the equation of state are also discussed.
\end{abstract}


\maketitle

\section{Introduction}

Theoretical methods developed in liquid-state physics\cite{HMc86}
have proven to be powerful tools to understand and predict the properties of polymer
liquids.\cite{SC97} An example is the polymer reference interaction
site model (PRISM) theory, which allows one to compute the pair
correlation functions and thermodynamic properties of amorphous
polymer liquids from knowledge of the polymer architecture. PRISM
theory is an extension to polymers by Curro and
Schweizer\cite{CS87,SC87} of the reference interaction site model
(RISM) theory developed by Chandler and
Andersen\cite{CA72,Chandler82} for small molecule liquids. The
theory can be applied to atomistic polymer models by computing the chain structure
 self-consistently by Monte Carlo (MC) simulation.\cite{SHC92} When implemented in this
manner, the theory is exact for intramolecular correlations at low density.

Comparisons of theoretical intermolecular radial distribution
functions and structure factors with atomistic molecular dynamics (MD) simulations of the
full, many-chain system\cite{CSGK89,HGC05} have shown that PRISM
theory gives a good qualitative, and sometimes semi-quantitative,
description of the structure of flexible, bead-spring polymer
liquids with strong repulsive interactions. Unfortunately, similar
comparisons\cite{CWGW99,HGC05} on more realistic polymer models
reveal that the theory becomes less quantitative as fixed rotational
angles, torsional potentials, overlapping interaction sites, and
softer potentials are incorporated into the models. This trend was
clearly seen in a study\cite{CWGW99} comparing
self-consistent PRISM theory with simulations on polyethylene (PE) melts.
The inaccuracy of PRISM theory can be traced to the approximations inherent in the use of the atomic Percus-Yevick (PY) closure,\cite{HMc86,SC97} which is currently considered the most accurate for liquids at melt densities with structure dominated by short-ranged repulsive interactions.

A computationally more demanding theory for molecular liquids was suggested by Laria, Wu and Chandler\cite{Laria91} and later derived by Donley, Curro, and McCoy (DCM)\cite{DCM94} using the classical density functional theory (DFT) of Chandler, McCoy, and Singer.\cite{CMS86a} In the theory, the expression for the radial distribution function consists of an average over the configurations of two molecules in a self-consistently determined field, i.e., a solvation potential. Thus the many-chain polymer liquid problem reduces to the much simpler one of two polymer chains. The derived solvation potential has a hypernetted-chain (HNC) form, and so this ``two-molecule'' equation can also be considered a closure to the RISM equation. This two-molecule theory was shown\cite{DCM94} to give more accurate predictions than RISM-PY theory for the pair correlations and equation of state of hard-core dimer liquids over a range of densities and bond lengths.

Subsequently, Yethiraj, Fynewever, and Shew\cite{YFS01} within a two-molecule theory derived a different expression for the solvation potential. This derivation employed a weighted DFT, which requires the equation of state of the liquid as an input.\cite{Wood94} They solved the two-molecule equation via MC simulation for a liquid of freely-jointed, and freely rotating, tangent sphere chains. They also found very good agreement for liquid structure between their theory and a full many-chain simulation. While requiring the equation of state to compute the properties of a homogeneous liquid lessens the predictive power of this weighted DFT theory, the good agreement does show the promising potential of two-molecule theories in general. 

A thread two-molecule theory, which fits within the framework of polymer self-consistent field theory, has recently also been developed.\cite{hu2017}

The purpose of the present investigation is to implement the version of two-molecule theory with an HNC-like solvation potential,\cite{Laria91,DCM94} hereinafter referred to as TM theory, to study PE liquids. A realistic chain model, commonly employed in molecular dynamics (MD) simulations, is utilized. The focus is to examine the accuracy of TM theory for the liquid structure and equation of state, particularly in comparison to state-of-the-art, self-consistent PRISM theory. 

The remainder of this paper begins with definitions of the system and correlation functions of interest. Then the theory for computing intramolecular correlations self-consistently within RISM and PRISM theory is discussed. Then the intermolecular portion of TM theory is described. This is followed by a description of the algorithm used to solve the theory numerically, though the discussion of the sampling methods employed in the one- and two-molecule MC simulations is deferred to the appendix. Then, results for PE liquids are presented and compared with PRISM-PY theory, and MD simulation data of full, many-chain liquids.

\section{Theory}
\subsection{Preliminaries}
Consider a liquid of $M$ identical polymer molecules in equilibrium in a volume $V$ at temperature $T$, giving a chain number density of $\rho=M/V$. Each chain is composed of $n_t$ types of spherical sites, with $N_k$ sites of type $k$. The number density of type-$k$ sites is then $\rho_k=N_k\rho$. Let the position of the $\alpha$th site of type $k$ on the $ith$ chain be denoted by $\textbf r_{\emph{i}k \alpha}$. Unless explicitly stated otherwise, in what follows all energies will be expressed in units of $\emph{k}_BT$, where $\emph{k}_B$ is the Boltzmann constant. 

For this system, the density-density correlation function between sites of type $k$ and $k'$ a distance $r\equiv \vert {\textbf r}\vert$ apart is defined as
\begin{equation}
  S_{kk'}(r) = \left\langle \left({\hat\rho}_{k}({\textbf r}) - \rho_k \right)
                             \left({\hat\rho}_{k'}({\textbf{0}}) - \rho_{k'} \right) \right\rangle,
\label{eq:Sofr}
\end{equation}
where
\begin{equation}
 {\hat \rho}_{k}(\textbf r) = \sum_{\emph{i},\alpha}
                             \delta \left(\textbf{r}-\textbf{r}_{\emph{i}k \alpha} \right)
\end{equation}
is the microscopic density of sites of type $k$ at point $\textbf{r}$, with $\delta(\textbf r)$ being the Dirac delta function, and the bracket denoting a thermodynamic average
\begin{equation}
  \Bigl\langle...\Bigr\rangle \equiv \dfrac {\int\prod_i^M\emph{d}\Re_i
         \cdots \exp\left[-V^{(M)}(\Re_1,\cdots,\Re_M)\right]}
         {\int\prod_i^M\emph{d}\Re_i
         \exp\left[-V^{(M)}(\Re_1,\cdots,\Re_M)\right]}.
\label{eq:thermoavg}
\end{equation}
Here, $\Re_\emph{i} = \{\textbf{r}_{i11},\textbf{r}_{i12},\cdots, \textbf{r}_{i21}, \cdots \}$ is the set of coordinates of the $\emph{i}th$ chain, and the potential, $V^{(M)}$, contains all the system interactions. The Fourier transform of Eq.~\eqref{eq:Sofr} is the partial structure factor, ${\hat S}_{kk'}(q)$, where $q\equiv\vert\bf q\vert$ is the wavevector conjugate to $r$. It is proportional to the intensity of scattered radiation in the single scattering limit, and thus can be used to make contact with experiment.

It is convenient to represent $S_{kk'}(r)$ as the sum of intra- and intermolecular contributions
\begin{equation}\label{S-def}
S_{kk'}(r) = \Omega_{kk'}(r) + H_{kk'}(r),
\end{equation}
where the intramolecular structure function
\begin{equation}
  \Omega_{kk'}(r) = \rho \sum_{\alpha,\beta}\left \langle
                  \delta(\textbf{r}-\textbf{r}_{1k \alpha}+\textbf{r}_{1k'\beta})
                  \right \rangle,
\label{omegadef}
\end{equation}
and the intermolecular correlation function 
\begin{equation}
  H_{kk'}(r)= \rho_k \rho_{k'}h_{kk'}(r) = \rho_k \rho_{k'}[g_{kk'}(r)-1],
\label{eq:Hr}
\end{equation}
with $g_{kk'}(r)$ being the radial distribution function 
\begin{equation}
  g_{kk'}(r)= \frac {1}{N_kN_{k'}} \sum_{\alpha,\beta}
           \left  \langle V\delta (\textbf{r} -\textbf{r}_{1k \alpha}
           +{\textbf{r}}_{2k'\beta})\right \rangle.
\label{eq:gofr}
\end{equation}

If the structure of the molecule depends on its environment, which is the case for flexible polymers, then the intramolecular structure function, $\Omega_{kk'}(r)$, and the intermolecular radial distribution function, $g_{kk'}(r)$, must be computed self-consistently. Methods for doing so within TM theory are described below.

\subsection{Intramolecular structure: one-molecule theory} 
It is instructive to examine first how the intramolecular structure function is obtained for a realistic polymer model. In (intramolecularly) self-consistent RISM\cite{CSR84} or PRISM\cite{SHC92} theory, $\Omega_{kk'}(r)$ is computed in a self-consistently determined field. This computation in PRISM theory is often done by a MC simulation over the conformations of a single molecule.\cite{HGC05} The field itself depends on $\Omega_{kk'}(r)$, but also on intermolecular correlations through $g_{kk'}(r)$. The theory is then usually completed by using PRISM theory, with a suitable closure, to compute $g_{kk'}(r)$ given $\Omega_{kk'}(r)$.

The total system potential energy, $V^{(M)}$, appearing in Eq.~\eqref{eq:thermoavg} can be written
as
\begin{equation}
  V^{(M)}(\Re_1,\cdots,\Re_M)=\sum_i^M U_0(\Re_i)+
                   \frac{1}{2}\sum_{i\neq j}^M U(\Re_i,\Re_j),
\label{VMeq}
\end{equation}
where $U_0(\Re)$ is the internal energy of a chain, and 
\begin{equation}
U(\Re_i,\Re_j)=\sum_{k,k'}\sum_{\alpha,\beta}
  u_{kk'}(\vert {\bf r}_{ik\alpha}-{\bf r}_{jk'\beta}\vert),
\label{Ueq}
\end{equation}
is the interaction energy between chains $i$ and $j$, with $u_{kk'}(r)$ being the potential between two sites 
of type $k$ and $k'$ a distance $r$ apart.

The intra-chain potential energy $U_0$ is a sum of bond and nonbonded interaction energies. The
bond energy consists of stretching, bending and torsion interactions between
neighboring sites on the chain. The possibly longer-ranged nonbonded interactions 
have the same origin as the intermolecular interactions between sites on different chains.

\begin{figure}
\includegraphics[scale=0.39,trim= 0.3in 0.3in 0.0in 0.4in]{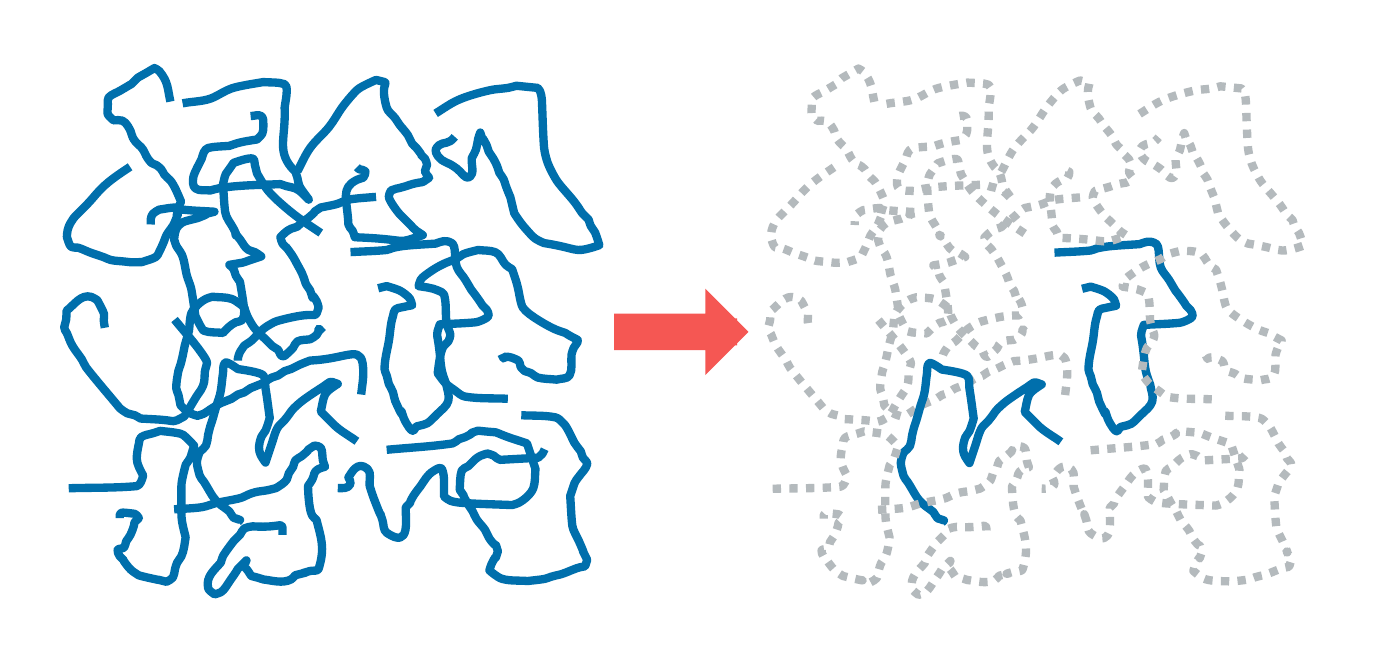}
\caption{\label{TMconcept} Pictorial representation of the two-molecule concept. A liquid of many molecules, illustrated here as identical, flexible and linear polymers, is modeled as two molecules in an external field that mimics the effects of the original surrounding molecules. }
\end{figure}

The essence of two-molecule theory is to realize that the exact expression for the radial distribution, Eq.~\eqref{eq:gofr}, can always be represented as an average over the configurations of two molecules, denoted as $\Re_1$ and $\Re_2$ here, in an effective field that itself depends on these configurations.\cite{CMS86b,DCM94} In other words, the degrees of freedom of the other molecules can always be integrated out first, at least in a formal sense. See Figure \ref{TMconcept}.

Similarly, one can always represent the expression for the intramolecular correlation function, Eq.~\eqref{omegadef}, as an average over the conformations of a single molecule in an effective field.\cite{Edwards1965,Melenkevitz93}  In that spirit, replace the potential between one chain and all others by a solvation potential $W(\Re)$, which depends only on the configuration of that one chain. With this, the intramolecular correlation function defined in Eq.~\eqref{omegadef} is now expressed as
\begin{equation}\label{omega-sol}
  \Omega_{kk'}(r)=\rho\sum_{\alpha,\beta}
            \left\langle\delta \bigl (\textbf r - \textbf r_{1k \alpha} +
            \textbf r_{1k'\beta}\bigr)\right\rangle_{\Re_1},
\end{equation}
where the term $\langle...\rangle_{\Re_1}$ represents an average over the configurations of
one chain,
\begin{equation}
  \Bigl\langle...\Bigr\rangle_{\Re_1} \equiv \int \emph{d}\Re_1P(\Re_1)
           \cdots
\end{equation}
with 
\begin{equation}
 P(\Re_1) = \frac {\exp[-V^{(1)}(\Re_1)]}
           {\int \emph{d}\Re_1 \exp[-V^{(1)}(\Re_1)]}
\label{p-distrib}
\end{equation}
being the probability density for chain 1 to be in configuration $\Re_1$, given an effective one-molecule
potential
\begin{equation}
  V^{(1)}(\Re_1)=U_0(\Re_1) + W(\Re_1).
\end{equation}

To make the theory tractable, $W(\Re_1)$ is usually approximated as pairwise decomposable
\begin{equation}
  W(\Re_1)\approx \sum_{k,k'}\sum_{\alpha,\beta}w_{kk'}
         (|\textbf r_{1k\alpha}-\textbf r_{1k'\beta}|),
\label{eq:pairwise}
\end{equation}
where $w_{kk'}(r)$ is the solvation, i.e., medium-induced, potential between sites of type $k$ and $k'$ a distance $r$ apart. 

Given this similarity to two-molecule theories, it is helpful to denote this manner of computing intramolecular correlations as ``one-molecule'' theory.

Using density functional techniques, Melenkevitz, Curro and Schweizer\cite{Melenkevitz93} derived an HNC form for $w_{kk'}(r)$:
\begin{equation}\label{W-HNC}
  {\textbf W}(r) = -{\textbf C}\ast {\textbf S}\ast {\textbf C}(r).
\end{equation}
Here, the boldface capital objects denote square matrices of rank $n_t$, and the
asterisks denote convolutions with implicit sums over common indices., e.g.,
\begin{equation}
    \textbf{A} * \textbf{B}(\textbf r) \equiv \sum_l 
    \int {{\emph{d}}\textbf {r}' A_{kl}(\textbf r') B_{lk'}(\textbf r'-\textbf r)}.
 \label{eq:convolution}
\end{equation}
In Eq.~\eqref{W-HNC}, matrix ${\textbf S}$ has components $S_{kk'}(r)$, which are given by Eq.~\eqref{eq:Sofr}. The matrix $\textbf{C}$ has components $c_{kk'}(r)$, which are direct correlation functions defined by the PRISM integral equation\cite{CS87,SC87}
\begin{equation}
\label{prism}
  \mathbf H(r) = \mbox{\boldmath $\Omega$} * \mathbf C * \mbox{\boldmath $\Omega$}(r)
       + \mbox{\boldmath $\Omega$} * \mathbf C *\mathbf  H(r).
\end{equation}
As is well known, this equation is derived from the RISM equation\cite{CA72,Chandler82} in the approximation that the direct correlation functions are dependent only on the type of sites, and for long polymers the chemical difference between the internal and end monomers is ignored.

As with all Ornstein-Zernike theories, Eq.~\eqref{prism} requires a closure to relate $\textbf{C}$ back to $\textbf {H}$ and $\mathbf {\Omega}$.\cite{HMc86} Unlike though the diagrammatically proper integral equation theory\cite{CSL82,Rossky1984} and the formally equivalent (at least in one case) optimized cluster theory,\cite{Lupkowski1987,Melenkevitz97} for which a closure can be derived diagrammatically, $\textbf{C}$ as defined by the RISM equation cannot be represented as a subset of diagrams of $\textbf{H}$. So the conventional method to derive a closure is not possible here. However, there have been many ansatz closures for Eq.~\eqref{prism}. In  particular, when the atomic, molecular or polymeric liquid is dominated by short-range interactions, the PY closure has shown to give good agreement with simulation.\cite{HMc86,SHC92,YS92,CSGK89,EK90} It is
\begin{equation}\label{PY-closure}
  c_{kk'}(r) = g_{kk'}(r)\left[1-\exp(u_{kk'}(r))\right].
\end{equation}
With this, the system of equations is completely specified and the
intermolecular correlation function can be determined by solving the PRISM equation. 

Self-consistent (one-molecule +) PRISM-PY theory has been applied to realistic polymer melts and blends, such as PE, polypropylene and poly-isobutylene.\cite{CWGW99,PCG01,HWCG03} In all applications, theory is in good
semi-quantitative agreement with many-chain, atomistic simulations. However the
approximation of the closure was also found to provide some
unphysical structures to the intermolecular correlation function.
The PY closure systematically underpredicts the degree of
intermolecular packing, and overpredicts the the compressibility of
polymer liquids.\cite{CWGW99} Furthermore, the closures do not work
well for systems with attractive tails,\cite{SC97} in which case the
structure factor will often become very large at low wavevector.

\subsection{Intermolecular structure: TM two-molecule theory}
An alternative to the diagrammatic approach is to derive a separate equation for $g_{kk'}(r)$, which itself involves the RISM direct correlation function in some way. DCM\cite{DCM94} made use of a molecular DFT\cite{CMS86a} to approximate the radial distribution function as
\begin{eqnarray}
  g_{kk'}(r) = &&\frac {1} {N_kN_{k'}}\sum_{\alpha,\beta}\Bigl\langle\Bigl\langle
   V\delta(\textbf r - \textbf r_{1k \alpha} + \textbf r_{2k'\beta}) \nonumber \\
   &&\times \exp(-V^{(2)}(\Re_1,\Re_2))\Bigr\rangle\Bigr\rangle_{\Re_1,\Re_2},
\label{gr-sol}
\end{eqnarray}
where the effective potential between molecules 1 and 2 has a pair-wise form,
\begin{equation}
V^{(2)}(\Re_1,\Re_2)= \sum_{k,k'}\sum_{\alpha,\beta} 
    u_{kk'}^{eff}(|\textbf r_{1k\alpha}-\textbf r_{2k'\beta}|),
\end{equation}
with 
\begin{equation}
u_{kk'}^{eff}(r) = u_{kk'}(r) + w_{kk'}(r),
\end{equation}
$w_{kk'}(r)$ being a solvation potential. The double brackets correspond to an average over the configurations of two chains with respect to their independent intramolecular probability densities:
\begin{equation}
  \Bigl\langle\Bigl\langle...\Bigr\rangle\Bigr\rangle_{\Re_1,\Re_2} \equiv 
     \int\prod_{i=1,2}\emph{d}\Re_i P(\Re_i)\cdots
\end{equation}
Thus the many-molecule average of Eq.~\eqref{eq:gofr} reduces to a two-molecule average.

Though a different approximation was used, the intermolecular solvation potential, $w_{kk'}(r)$, derived by DCM has the same HNC form, Eq.~\eqref{W-HNC}, as the intramolecular one derived by Melenkevitz et al.\cite{Melenkevitz93}

The method of DCM has been extended to compute possibly more accurate solvation potentials,\cite{Donley05b,Tanaka14} but the predictions of those are left for future work. Other forms of the solvation potential have been suggested,\cite{Grayce94,DCM94} but these are generalizations of atomic ones that have not been shown to perform well for charged molecules.  The effect of these latter potentials for PE has been examined elsewhere.\cite{HuiminThesis} 

Given the form of Eq.~\eqref{gr-sol}, a possible advantage of TM theory is that it allows a natural way to handle potentials of a wider variety, including those with attractive tails, which can be problematic for other closures to RISM or PRISM theory. For example, without some modification, no other closure to the PRISM equation has been shown to work consistently well for liquids dominated by Coulomb interactions.\cite{Donley05a,Donley1998}

\section{\label{sec:polychainmodel}Polyethylene Chain Model}
In this work, TM theory is applied to a PE melt, which is the simplest realistic polymer system. The PE homopolymer is modeled as a linear chain of overlapping, spherically symmetric sites.  Each
site is a united atom consisting of one carbon and two bonded hydrogen atoms, ${\rm CH}_2$.
The chemical difference between ${\rm CH}_2$ and the two terminal ${\rm CH}_3$ groups is neglected. In this manner, the site labeling of correlation functions such as $g(r)$ will be dropped for the remainder of this work.  Also, let $N_m$ and $\rho_m$ denote the number of  $\rm CH_2$ monomers per chain (degree of polymerization) and monomer density, respectively.  

The nonbonded interactions between sites are modeled by the
Lennard-Jones (LJ) potential. In this work, both the full potential with an attractive tail, and a purely repulsive form will be considered. The full potential is
\begin{equation}
  u_{f}(r)= 4\epsilon \left[\left( \frac{\sigma }{r}\right) ^{12}-
      \left( \frac{\sigma }{r}\right) ^{6}\right],
\label{fullLJ}
\end{equation}
where $\epsilon$ and $\sigma$ are parameters fixing the energy and
length scale, respectively. For simplicity, this full potential was implemented with a cut-off distance, $r_{cut}$, at which it went to zero. Here, $r_{cut}=10\sigma$. Setting it larger changed the results by only a negligible amount. Following Weeks, Chandler and Andersen,\cite{WCA71} the purely repulsive potential has the form
\begin{equation}
  u_{r}(r)=\left\{ \begin{array}{cc}
      u_f(r) + \epsilon,
      & r\leq 2^{1/6}\sigma \\
      0, & r>2^{1/6}\sigma
\end{array}\right .
\label{repulLJ}
\end{equation}

As in the earlier studies\cite{CWGW99,PCG01,HWCG03} on polymer
liquids, the bond length, $l$, between sites on the chain is fixed, so there is no stretching potential. The bond angle $\theta$
between three nearest sites varies in a harmonic bending potential
about a fixed angle $\theta_b$,
\begin{equation}
  V_b(\theta)=\frac {k_b}{2}(\theta-\theta_b)^2.
\end{equation}
The dihedral angles $\phi$ between four nearest sites are subjected
to the OPLS torsional potential
\begin{equation}
  V_t(\phi)=\sum_{i=0}^3a_i\cos^i(\phi),
\end{equation}
where $\cos(\phi)$ can be expressed in terms of the positions of the four sites.\cite{PCG01}

Lastly, nonbonded potentials, both the Lennard-Jones and solvation, are restricted intramolecularly to apply only between site pairs that are separated by a distance along the chain backbone of four sites or more.\cite{CWGW99} This restriction is standard in MD and MC simulations for the Lennard-Jones as the bond stretching, bending and torsional potentials are assumed to model completely the interactions between 1-2, 1-3 and 1-4 site pairs.\cite{Plimpton1995} The solvation potential, since it mimics the effects of the surrounding chains, should in principle not be limited in this way though. However, one-molecule theory, as is traditionally done in PRISM theory, approximates the solvation potential as dependent only on site type, not site index. Intramolecularly, this approximation does not account completely for the qualitative difference in screening between sites of varying distance along the chain backbone.\cite{deGennes} For long chains of PE at melt densities, at short backbone distances, the chain has a stiff, self-avoiding structure, being largely unaffected by its environment, but at long distances the structure is ideal. The above restriction of the solvation potential intramolecularly to the same site pairs as the Lennard-Jones is then the least one, the size of propane (C4H10) being independent of density.\cite{Yaws2014}

The parameters associated with the interaction potentials for PE,  which were obtained from
Siepmann et al.,\cite{SKS93,SKS95} are given in Table \ref{tabPOT}. These parameters were also used
in the MD simulations of Curro et al.\cite{CWGW99} and one by us, which allows a direct comparison between theory and simulation.
\begin{table}
\caption{Potential parameters for PE.}
\begin{tabular}{ccc}
\hline \hline
$\rm Parameter$ & $\rm Value$ & $\rm Units$  \\
\hline
$l$ & 1.54 & \AA  \\
$k_b$ & 124.18 & $\ {\rm kcal/(mol\mbox{-}rad^2)}$  \\
$\theta_b$ & 114.0 & $\rm deg$ \\
$a_0$ & 2.007 & ${\rm kcal/mol}$ \\
$a_1$ & -4.012 & \\
$a_2$ & 0.271 & \\
$a_3$ & 6.290 & \\
$\epsilon/k_B$ & 47.0 & deg K \\
$\sigma$ & 3.93 & \AA \\
\hline\hline 
\end{tabular}
\label{tabPOT} 
\end{table}

\section{\label{sec:numerical}Numerical solution}
In this section an overview is presented of the numerical schemes for solving both the PRISM-PY and TM theories. More details about the numerical calculations for self-consistent (one-molecule +) PRISM-PY theory are available in P\"{u}tz et al.\cite{PCG01} and Heine et al.\cite{HWCG03} The computer code used here was an extension of that created for the research described in P\"{u}tz et al.\cite{PCG01}

First, all functions of $r$ and reciprocal space wavevector $q$ were solved on grids of $N_r$= 2048 points with a spacing of $\Delta r =0.1$ \AA, and $\Delta q =\pi/(N_r\Delta r)$, respectively. All convolutions, such as the one in Eq.~\eqref{eq:convolution}, were computed by Fourier transforms, with the FFTW software library\cite{fftw3} being used.

The approach to solve for $g(r)$ and $\Omega(r)$ self-consistently was as follows. Initially a guess was made for the solvation potential, $W(r)$, and a single chain MC simulation, discussed in the Appendix, was performed using this solvation potential to obtain $\Omega(r)$ from Eq.~\eqref{omega-sol}. With $\Omega(r)$ and an initial guess for $\gamma(r) = h(r) - c(r)$, Eqs.~\eqref{PY-closure} and \eqref{prism} were then solved for $c(r)$ and $h(r)$, respectively, by iteration. The modified method of direct inversion in iterative subspace (MDIIS) of Kovalenko, Ten-no and Hirata\cite{Kovalenko1999,Ishizuka2012} was used to obtain a new estimate for $\gamma(r)$ for each iteration, with a mixing ratio of the old to new guess of 4:1 typically, though sometimes it was set to 49:1 when the initial guess was poor. The error on $\gamma(r)$ for the $i$th iteration was estimated as
\begin{equation}
Err_\gamma^{(i)} = \int d{\bf r} \left[\Delta \gamma^{(i)}(r)\right]^2
\label{gammaerror}
\end{equation}
where $\Delta \gamma^{(i)}(r) = \gamma_{out}^{(i)}(r) - \gamma_{in}^{(i)}(r)$, i.e., the difference between the input and output values for this iteration.
The error threshold for convergence in the MDIIS scheme was set to $10^{-6}$. Depending on the quality of the initial guess, it took up to 30 iterations to achieve convergence, with an average of about 7. 

Once $\gamma(r)$ had converged, a new estimate of the solvation potential $w(r)$ was obtained  from Eq.~\eqref{W-HNC} using $c(r)$, $h(r)$ and the last estimate of $\Omega(r)$ within MDIIS with a mixing ratio of 4:1, old to new. The error for $w(r)$ was estimated in the same way as for $\gamma(r)$ in Eq.~\eqref{gammaerror},  but with the error threshold set to $10^{-5}$. It took up to 35 iterations to achieve convergence, with an average of about 20.
 
In order to reduce computation time, a MC reweighting
scheme\cite{HWCG03,PCG01,SF88} was utilized in the self-consistent
procedure to reuse the $N_s$ conformations, $\{\Re_{old}\}$, generated with a previous, i.e., old,
value of the total solvation energy, $W(\Re)$, given by Eq.~\eqref{eq:pairwise}. Let ${\tilde\Omega}_{old}^{(j)}(r)$ be the contribution to $\Omega(r)$ from an old conformation state $j$, $\Re_{old}^{(j)}$. Then the new intramolecular correlation function was computed as
\begin{eqnarray}
 \Omega_{new}(r)
  =&&\frac{1}{Z}\sum _{j=1}^{N_s}{\tilde\Omega}_{old}^{(j)}(r) \nonumber \\
 \times && \exp\left [-W_{new}(\Re_{old}^{(j)}) + W_{old}(\Re_{old}^{(j)})\right] 
\end{eqnarray}
where
\begin{equation}
   Z=\sum _{j=1}^{N_s}\exp\left [-W_{new}(\Re_{old}^{(j)}) + W_{old}(\Re_{old}^{(j)})\right] ,
\end{equation}
and $W_{new}$ and $W_{old}$ refer to the new and old value of the total solvation energy, respectively.
The statistical reliability of the reweighting scheme was determined
by applying the empirical criterion\cite{PCG01}
\begin{equation}
  \min \left\{ {\tilde Z},1/{\tilde Z}\right\} >\zeta,
\end{equation}
where ${\tilde Z} = Z/N_s$ is the average Boltzmann reweight for a state. The value of $\zeta$ was set to obtain at least two iterations without reweighting, and varied between 0.25 to 0.9, the smaller value being for the largest chain lengths.  If the difference between the new and old solvation potentials were large
enough for the criterion to fail, then a new set of conformations was
generated.

To solve the TM theory, the same scheme as above to determine $\Omega(r)$ via MC simulation and iteration on $w(r)$ was used. To solve for $h(r)$, $c(r)$ instead of $\gamma(r)$ was iterated to achieve convergence. Given $\Omega(r)$ and an initial estimate for $c(r)$, the PRISM equation, Eq.~\eqref{prism}, was solved for $h(r)$, denoted here as $h_{prism}(r)$. Then with $\Omega(r)$, $c(r)$ and $h_{prism}(r)$, a new estimate for $w(r)$ was determined using Eq.~\eqref{W-HNC}. This solvation potential, along with a set of single chain conformations, $\{\Re\}$, generated from the one-molecule simulation, was then used in the TM equation for $g(r)$,  Eq.~\eqref{gr-sol}, to obtain a new estimate for $h(r)$, denoted here as $h_{tm}(r)$. A direct sampling scheme, described in the Appendix, was used to evaluate Eq.~\eqref{gr-sol} by simulation. The change in $c(r)$ in the iteration step was approximated in a quasi-Newton-Raphson way as\cite{DCM94,Donley2014}
\begin{equation}
\Delta c(r) = \rho_m^2\Omega^{-1}*\Delta h*\Omega^{-1}(r)
\label{delCr}
\end{equation}
where $\Delta h = h_{tm}(r) - h_{prism}(r)$, and $\Omega^{-1}(r)$ is the functional inverse of $\Omega(r)$. This change in $c(r)$ was used as an input to the MDIIS algorthim to construct a new guess for $c(r)$, with a mixing of old to new of 4:1 typically. Likewise, the error on $c(r)$ was estimated in the same manner as for $\gamma(r)$ in Eq.~\eqref{gammaerror}, with the error threshold for convergence set to $10^{-6}$. It took up to 80 iterations to achieve convergence, with an average of about 10. Once $c(r)$ had converged, a new estimate for $w(r)$ was obtained for the next iteration on $\Omega(r)$. A diagram of the flow of this algorithm is shown in Figure \ref{figTCSolutionFlow} below.
\begin{figure}
\includegraphics[scale=0.61,trim= 0.1in 0.26in 0.10in 0.0in]{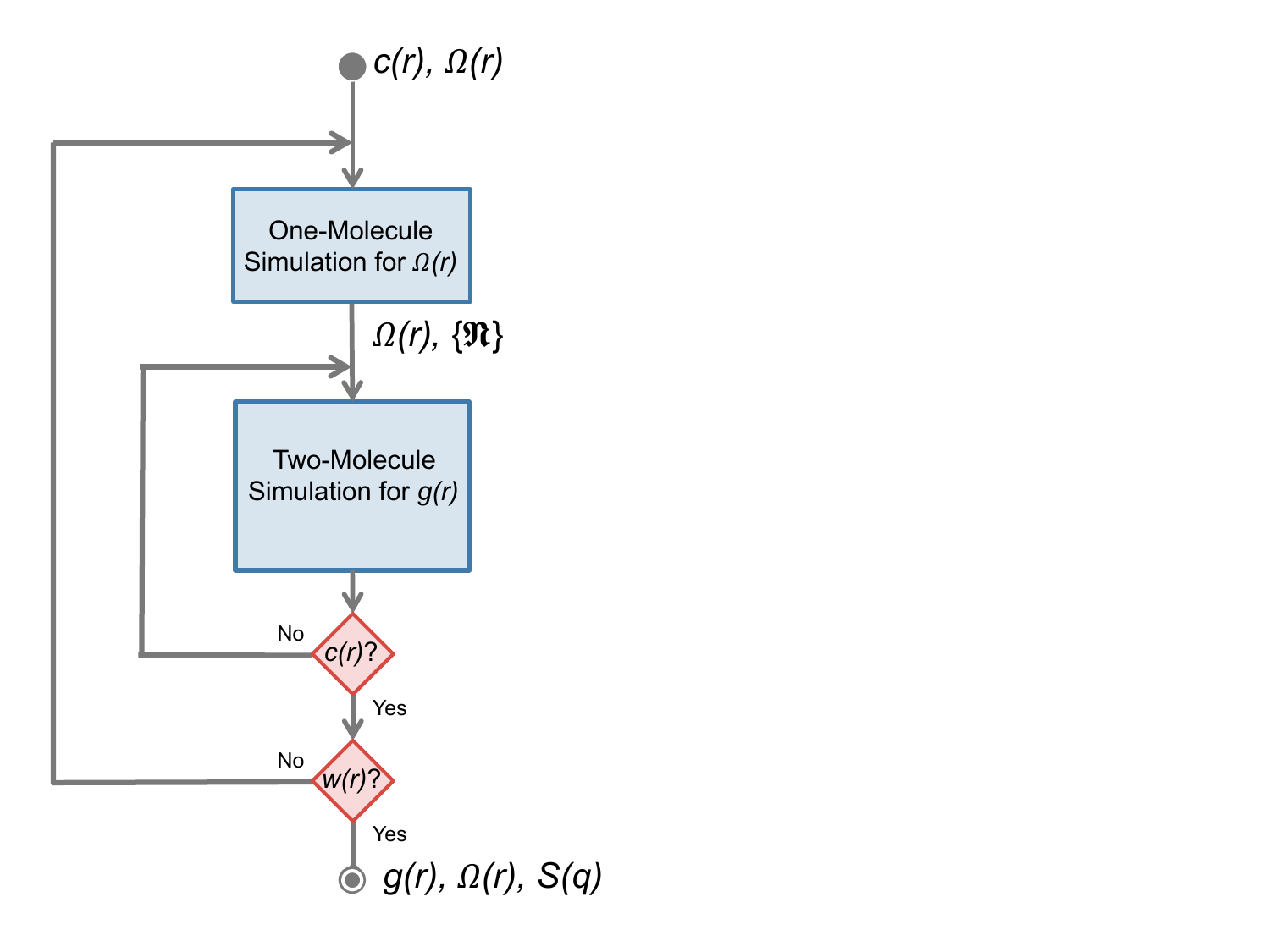}
\caption{Flow diagram of the algorithm, discussed in Sec. \ref{sec:numerical}, used to solve the TM theory. A red diamond denotes a test for convergence of, and iteration on, the given quantity. Inputs are initial guesses for the intramolecular correlation function $\Omega(r)$ and the direct correlation function $c(r)$, from which an initial guess for the solvation potential $w(r)$ is obtained. Outputs of the one-molecule simulation are a tentative value for $\Omega(r)$, and a set of chain conformations $\{\Re\}$. Indirect outputs of the two-molecule simulation are new values for $c(r)$ and thus $w(r)$. Final outputs are $\Omega(r)$, the radial distribution function $g(r)$, and the structure factor ${\hat S}(q)$.
}
\label{figTCSolutionFlow}
\end{figure}

\section{Results}
In this section, results are presented for the PE model discussed in Sec.~\ref{sec:polychainmodel} above. Predictions of the self-consistent (one-molecule +) PRISM-PY and TM theories are compared with data from atomistic, many-chain MD simulations of Curro et al.\cite{CWGW99} and us, which used the same PE model. Thermodynamic cases examined here will be from Curro et al.,\cite{CWGW99} and they are given in Table \ref{tabCases} below.
\begin{table}[h]
\caption{Thermodynamic cases for PE examined in this work.}
\begin{tabular}{llccc}
\hline \hline
Name & Potential & $\ N_m\ $ & $\rho_m\ (\text{\AA}^{-3})$ & $\ T$(deg K)   \\
\hline
PE24  & Full LJ     & 24 &  0.03123 & 405 \\
PE24r & Repulsive LJ & 24 &  0.03104 & 405 \\
PE66  & Full LJ    & 66 &   0.03294 & 448 \\
PE66r & Repulsive LJ & 66 &  0.03294 & 448 \\
\hline\hline
\end{tabular}
\label{tabCases}
\end{table}
We repeated the simulation for case PE24 with 1200 chains, but extended the run time to 100 $ns$, the LAMMPS simulation package\cite{Plimpton1995} being used.

\begin{figure}
\includegraphics[scale=0.453,trim= 0.3in 0.7in 0.0in 0.4in]{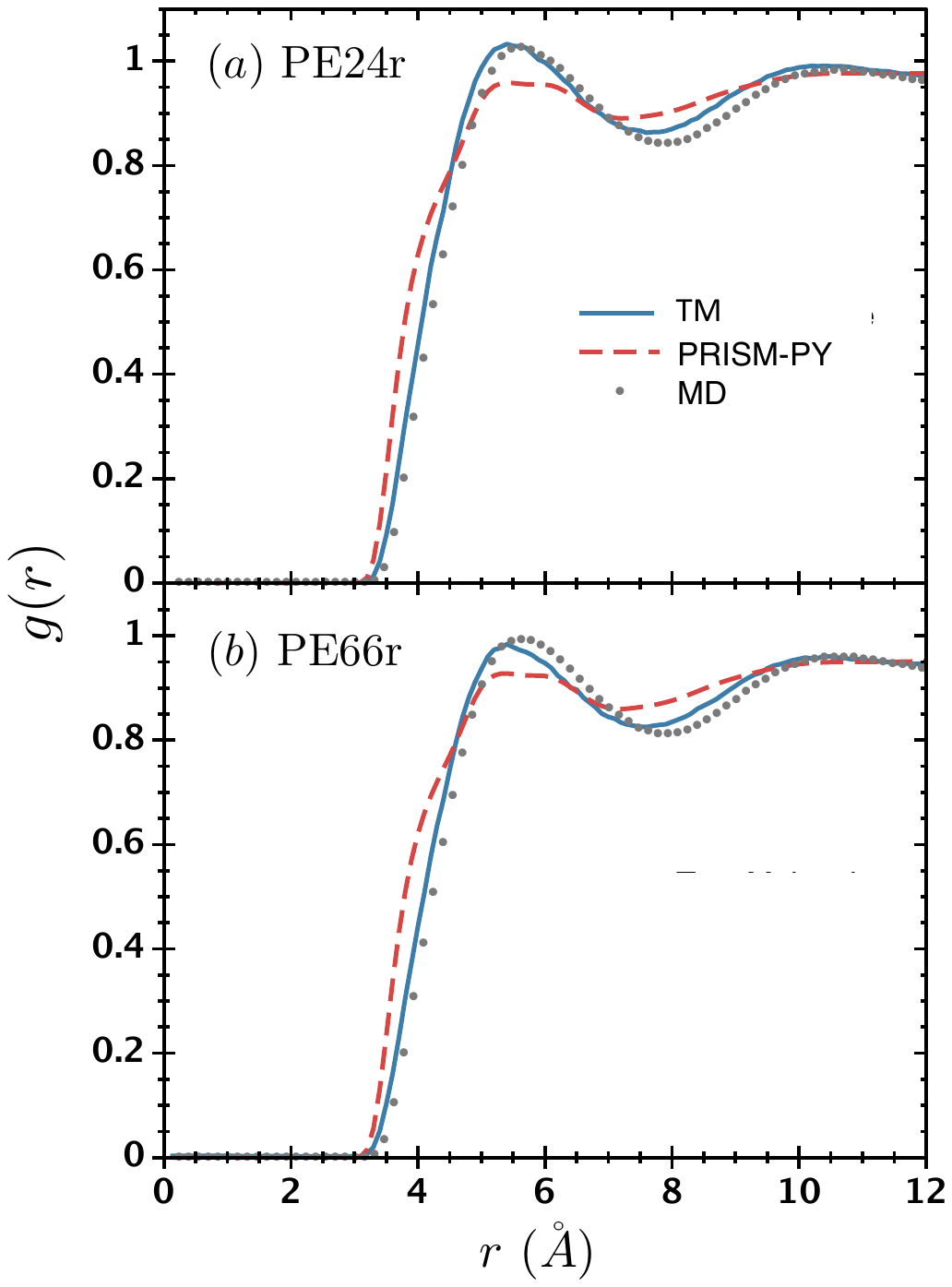}
\caption{\label{figgr} (a) Radial distribution function $g(r)$ as a function of radial distance $r$ for a PE melt of $N_m=24$ monomers interacting with repulsive Lennard-Jones potentials, i.e., case PE24r.The solid blue and dashed red lines correspond to results for TM and PRISM-PY theory, respectively. The grey dots are many-chain MD simulation data.\cite{CWGW99} (b) the same except the chain length $N_m=66$, i.e., case PE66r. }
\end{figure}

Figure \ref{figgr} (a) shows results for the site-averaged radial distribution function, $g(r)$, as a function of radial distance $r$ for case PE24r. As can be seen, there is noticeable improvement in the prediction for local structure with TM theory compared with PRISM-PY theory.  The latter is known to work well with coarse-grained chain models, but becomes less accurate with more realistic models.\cite{CWGW99} In particular, PRISM-PY theory tends to predict that intermolecular sites are closer together than found in the MD simulations. Moreover PRISM-PY theory predicts less local ordering since the height of the first peak and depth of the valley are less than the MD data. On the other hand, since TM theory models local packing exactly between pairs of molecules, its better agreement with the MD data is expected. The oscillation of the solvation shell ordering has a slightly shorter wavelength with TM theory compared with the MD data. One possible cause is the HNC form of the solvation potential, Eq.~\eqref{W-HNC}, which can lead to an overestimation of the strength of compression of the liquid on the two chains. The agreement between TM theory and the MD data is about the same for case PE66r, shown in subfigure (b), as PE24r, possibly indicating that the pair-wise approximation of Eq.~\eqref{eq:pairwise} is also performing well for longer chains.\cite{Melenkevitz93,Grayce94,DCM94,HuiminThesis}  

\begin{figure}
\includegraphics[scale=0.57,trim= 0.00in 0.0in 0.0in 0.0in]{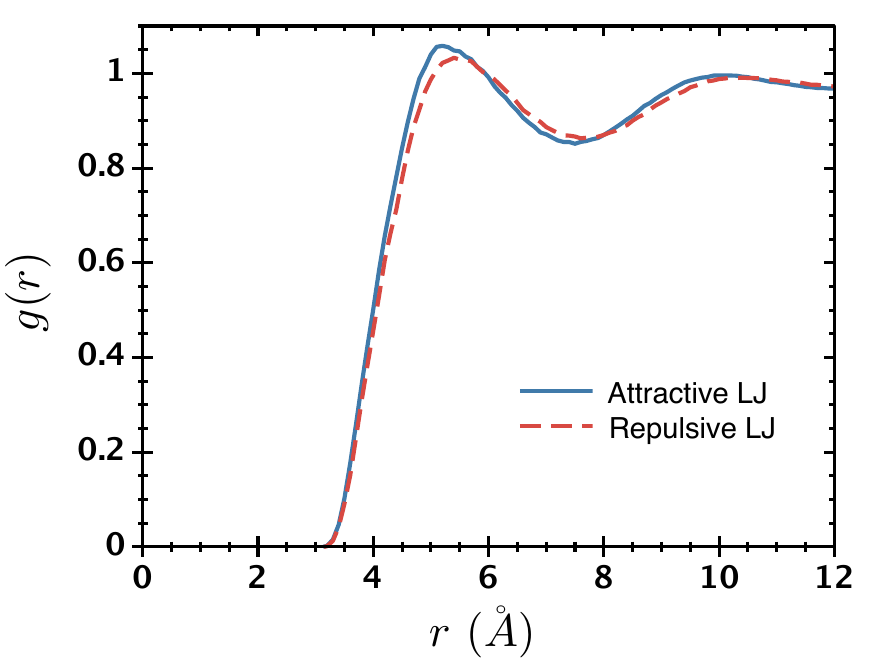}
\caption{\label{figgrAttrVsRepul} TM theory predictions for the radial distribution function $g(r)$ as a function of radial distance $r$ for a PE melt of $N_m=24$ monomers interacting with attractive and repulsive Lennard-Jones potentials.The solid blue and dashed red lines correspond to results for cases PE24 and PE24r, respectively.}
\end{figure}

Figure \ref{figgrAttrVsRepul} compares the radial distribution function in a liquid with attractive interactions, case PE24, to those with repulsive interactions, case PE24r. As can be seen, $g(r)$ does not change all that much when short-range attractive interactions are turned on, which is well-known behavior. 

\begin{figure}
\includegraphics[scale=0.71,trim= 0.1in 0.15in 0.0in 0.25in]{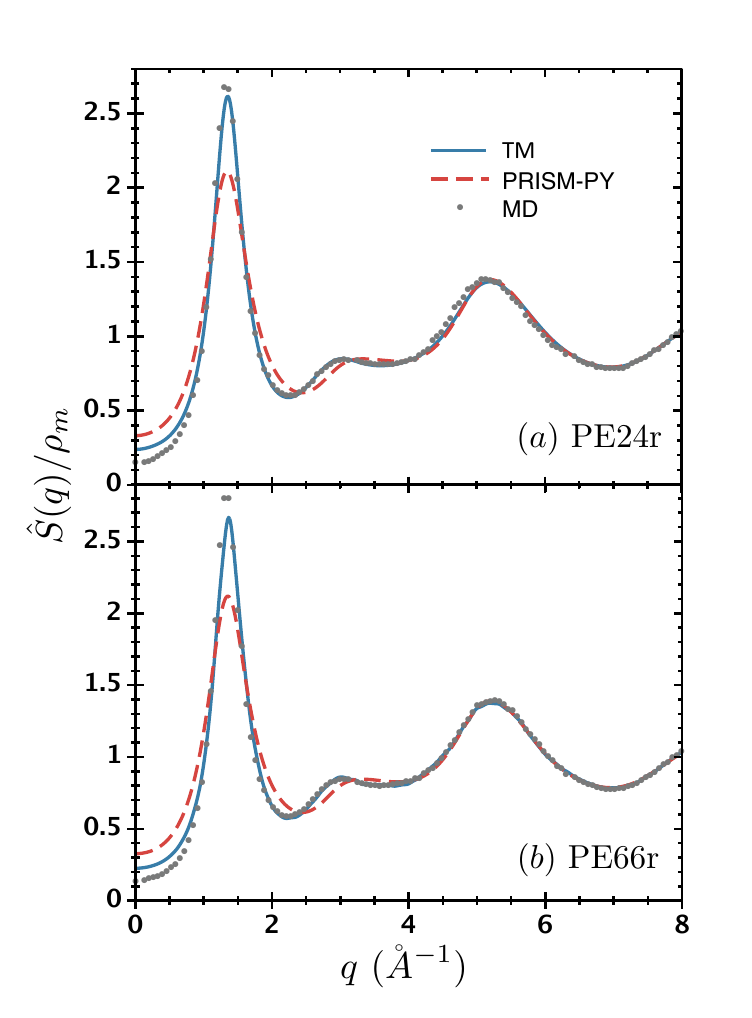}
\caption{\label{figSkRepulLJ} (a) Scaled structure factor ${\hat S}(q)/\rho_m$ as a function of wavevector $q$  for a PE melt of $N_m=24$ monomers interacting with repulsive Lennard-Jones potentials, i.e., case PE24r. The solid blue and dashed red lines correspond to results for TM and PRISM-PY theory, respectively. The grey dots are MD simulation data.\cite{CWGW99} (b) the same except $N_m=66$, i.e., case PE66r.}
\end{figure}

\begin{figure}
\includegraphics[scale=0.455,trim= 0.2in 0.7in 0.0in 0.4in]{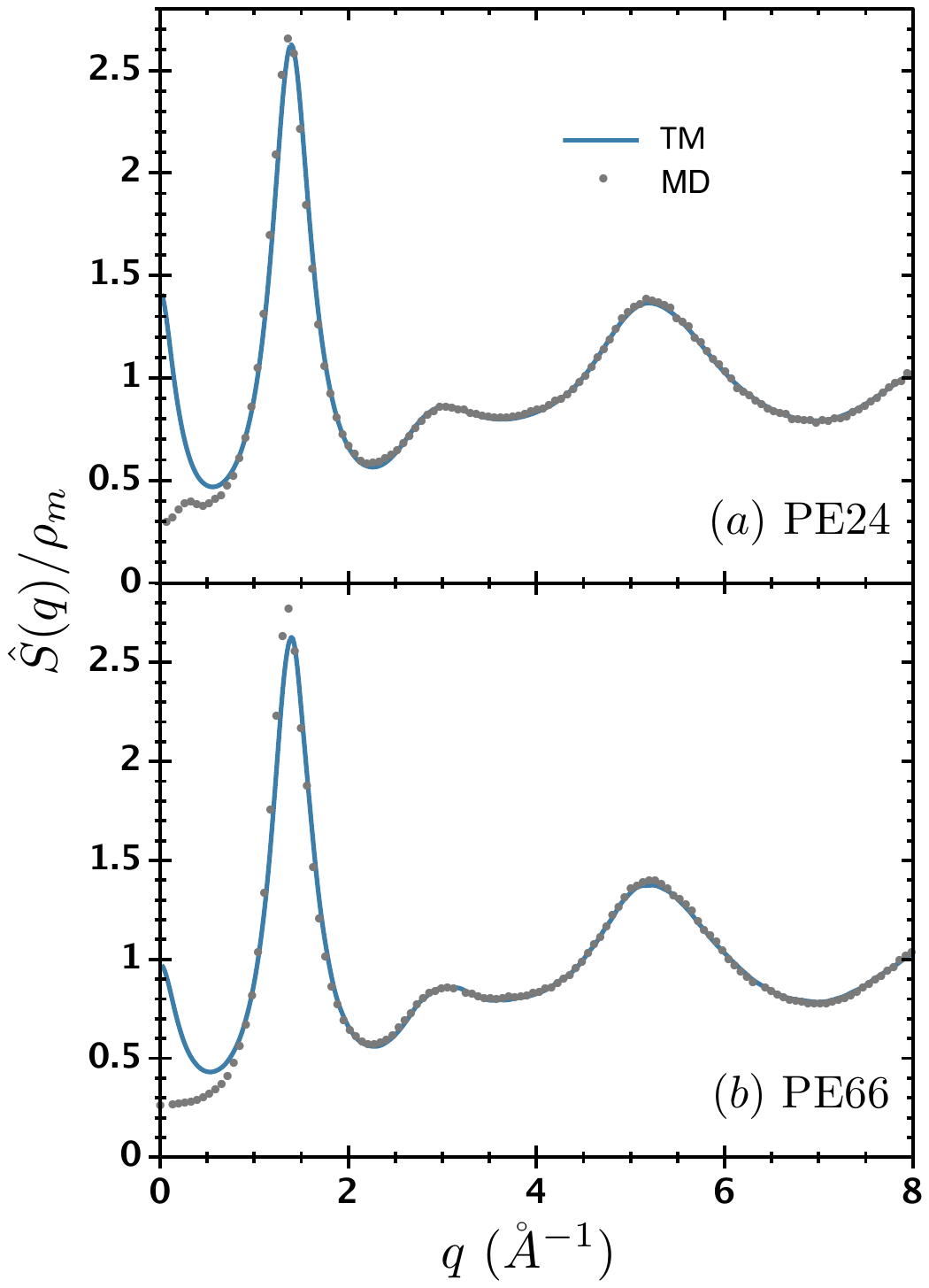}
\caption{\label{figSkAttrLJ} (a) Scaled structure factor ${\hat S}(q)/\rho_m$ as a function of wavevector $q$  for a PE melt of $N_m=24$ monomers interacting with full Lennard-Jones potentials, i.e., case PE24. The solid blue line is the prediction of TM theory, and the grey dots are MD simulation data from us. (b) the same except $N_m=66$, i.e., case PE66, and the MD simulation data are from Curro et al.\cite{CWGW99}}
\end{figure}

To examine longer wavelength properties, Figures \ref{figSkRepulLJ}  and \ref{figSkAttrLJ} show results for the scaled total structure factor ${\hat S}(q)/\rho_m$ as a function of wavevector $q$ for liquids interacting with repulsive and full Lennard-Jones potentials, respectively.  As can be seen in Figure \ref{figSkRepulLJ}, the agreement of TM theory with the MD data at intermediate to low $q$ is a definite improvement over that of PRISM-PY theory. The agreement does not seem to lessen for the longer chain length, case PE66r, shown in subfigure (b), which is encouraging. 
The TM and PRISM-PY theory predictions, and MD simulation data for ${\hat S}(0)/\rho_m$ are summarized in Table \ref{tabRg} below. This quantity is related to the isothermal compressibility $\kappa_T$ as
\begin{equation}
\kappa_T = {1\over \rho_m}\left ({\partial \rho_m \over \partial P}\right)_T 
                  =  \frac{{\hat S}(0)}{\rho_m^2},
\label{kappaT}
\end{equation}
where $P$ is the pressure. As can be seen, for the repulsive LJ liquid, the PRISM-PY prediction for ${\hat S}(0)/\rho_m$ is about 140\% times larger than the MD value, while the TM prediction is about 65\% times larger. In this way, TM theory predicts a less compressible liquid, in agreement with the MD data. 

On the other hand, for the full LJ liquid, TM theory results shown in Figure \ref{figSkAttrLJ} and Table \ref{tabRg} predict a much larger compressibility than measured by the MD simulation. The TM prediction for ${\hat S}(0)/\rho_m$ is five and four times larger than the MD values for cases PE24 and PE66, respectively, with ${\hat S}(q)$ showing an upturn at low $q$. Note though that for case PE24 the simulation data show a local maximum at $q\approx 0.3~\AA^{-1}$. These enhanced fluctuations are most likely the beginning of an upturn at low $q$, but because of the finite box size, fluctuations at smaller $q$ are suppressed. The correct value for ${\hat S}(0)/\rho_m$ then lies most likely between the TM and MD predictions. This overestimation of ${\hat S}(0)$ and thus the compressibility is possibly due to the known issue\cite{HMc86} of the HNC solvation potential, Eq.\eqref{W-HNC}, being too strong at high density, particularly for attractive Lennard-Jones liquids. Adding a ``bridge" correction to the solvation potential may be a remedy then.\cite{Tanaka14,Lee2011}
\begin{table}
\caption{Root mean square radius of gyration, $R_g$, and scaled zero wavevector susceptibility, $\ {\hat S}(0)/\rho_m$,  for the thermodynamic cases specified in Table \ref{tabCases} above. The MD data are from Curro et al.,\cite{CWGW99} except for PE24 which is from us.}
\begin{tabular}{ccccc}
\hline \hline
Case & Source & $R_g$(\AA)   &   $\ {\hat S}(0)/\rho_m$    \\
\hline
PE24r & TM & 6.50 & 0.239 \\
  & PRISM-PY & 6.61 & 0.331 \\
& MD & 6.70 & 0.150 \\
&&& \\
PE66r & TM & 12.57 & 0.223 \\
  & PRISM-PY & 13.17& 0.323 \\
& MD & 13.35 & 0.133 \\
&&& \\
PE24 & TM & 6.48 & 1.42 \\
  & MD & 6.72 & 0.289 \\
&&& \\
PE66  & TM & 12.48 & 0.98 \\
 & MD & 13.34 & 0.261 \\
\hline\hline
\end{tabular}
\label{tabRg} 
\end{table}

\begin{figure}
\includegraphics[scale=0.56,trim= 0.0in 0.1in 0.0in 0.0in]{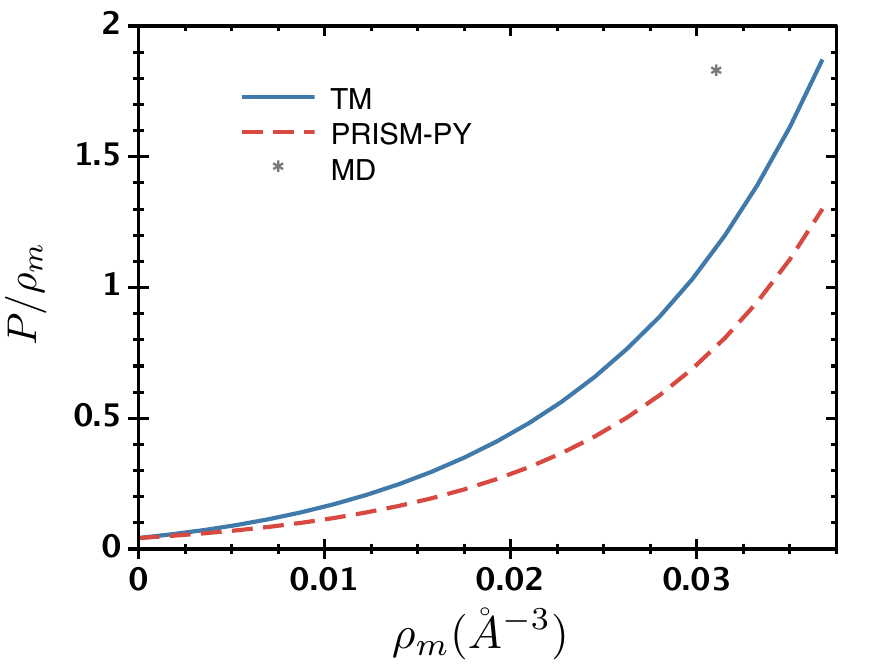}
\caption{\label{figPress} Scaled pressure $P/\rho_m$ ($k_BT=1$ here) as a function of monomer density $\rho_m$ along an isotherm at $T=405~K$ for a PE melt of $N_m=24$ monomers interacting with repulsive Lennard-Jones potentials. The solid blue and dashed red lines correspond to predictions of TM and PRISM-PY theory, respectively, computed from the compressibility route. The gray dot is from the MD simulation for case PE24r. }
\end{figure}

The pressure can be computed by integrating the inverse of the compressibility, Eq.~\eqref{kappaT}, with respect to the density. Figure \ref{figPress} shows the scaled pressure, $P/\rho_m$, as a function of monomer density, $\rho_m$, along an isotherm at $T=405~K$ for PE with $N_m=24$ and repulsive LJ interactions. As mentioned above, TM theory is exact at low density, so can be trusted there. For higher density, the sole simulation datum was obtained from the MD run for case PE24r. At this density, the TM and PRISM-PY predictions are 36 and 60\% lower than the MD pressure, respectively. These differences are consistent with those for the susceptibility, ${\hat S}(0)$, given in Table~\ref{tabRg}. Given past comparisons,\cite{DCM94} it is expected that the predictions of TM theory for the pressure from the virial route\cite{Honnell1987} would be more accurate.
 
As for single chain structure, Table \ref{tabRg} shows the chain radius of gyration, $R_g$, from the MD simulations and TM theory for all thermodynamic cases in Table \ref{tabCases} above, and PRISM-PY theory for cases PE24r and PE66r. Both the self-consistent (one-molecule +) PRISM-PY and TM theory predictions are lower than the MD values in all instances, TM theory being slightly less accurate for the larger chain length. Previous studies have suggested that an HNC solvation potential overestimates the force on a polymer chain\cite{Melenkevitz93} or a hydrated electron ring\cite{Laria91,Donley2014} at liquid densities, in some instances predicting an unphysical collapse. Heuristic remedies that weaken the strength of the solvation potential have been offered,\cite{Grayce94,DCM94,HuiminThesis,Donley2014} and theoretical attempts to incorporate effects beyond those contained in the HNC form have been made.\cite{Grayce94,Tanaka14} 

This difference between theory and MD simulation for $R_g$ affects the chain structure factor ${\hat\Omega}(q)$ only near $q=0$ though as agreement between the MD data of Curro et al. and self-consistent PRISM-PY theory at larger wavevectors, $q > 1\ \AA^{-1}$, has already been found.\cite{CWGW99} As expected, TM theory generated an ${\hat\Omega}(q)$ virtually identical to PRISM-PY theory and the MD simulation data at these larger wavevectors. This agreement between theory and simulation indicates that the HNC solvation potential is able to mimic the global effect of surrounding chains in an accurate way, at least when the difference in overall chain size is not too great.

\section{Summary and Discussion}
In summary, results have been presented for the first implementation of TM theory, i.e., two-molecule theory with an HNC-like solvation potential,\cite{Laria91,DCM94} for large molecules, in this case polymer melts of linear polyethylene. In comparison with many-chain MD simulation data, it was shown that the theory is a clear improvement on PRISM-PY theory for these realistic polymers, including predicting local structure and  the handling of the full Lennard-Jones interaction with its attractive tail. TM theory was solved mostly by a simple simulation, it employing an improved direct sampling technique to speed the computation of the radial distribution function via averaging over the configurations of the two chains. The equation of state for liquids with repulsive LJ interactions was also examined via the compressibility route. For the sole simulation data point at high density, it was found that both TM and PRISM-PY theory underestimate the pressure, but the former less so. 

Any differences between the predictions of TM theory and simulation seem to be due primarily to the overestimation of the strength of the solvation potential in the theory at melt densities, which with further research should be improved. The overall concept of modeling the structure of a liquid by two molecules in a self-consistent field appears to be a successful one.

Apart from providing an interesting, quantitative view of molecular liquids, another motivation for using the TM theory is its speed of computation compared with atomistic, many-chain MD or MC simulations. As discussed in the Appendix, the time to achieve convergence using the TM theory scales as the width of the box, while MD computations for melts scale at best as the width cubed. For case PE24, described in Table \ref{tabCases}, the MD run took 13 hours on 128 processors to reach 40 $ns$ of simulation time, which was sufficient to attain equilibrium and get good statistics. Rescaling this time to the same box size and number of processors as used in the TM theory calculation, the latter was 4200 times faster than the MD run, an improvement. Needless to say, no claim is made that the lower level algorithms used to implement the TM theory were optimized to any great degree, so further improvements in speed should be possible there. 

Finally, for development of technologically important materials,\cite{SC97} a goal of TM theory is to model realistically liquids with multiple types of molecules, such as alloys, and those with molecules of varying chemical architecture, such as branched polymers or copolymers. It is straightforward to generalize the algorithms presented in this work to multicomponent liquids for which each molecule is modeled with only one type of site, such as a polyelectrolyte solution with explicit counterions and implicit water. However, in TM theory it is known that for a liquid of heterogeneous molecules, such as carbon dioxide, that the Fourier transform of the multi-site analog of Eq.\eqref{delCr} can diverge as the wavevector $q\rightarrow 0$. This behavior can prevent the theory from producing a physically sensible numerical solution, if it converges at all. It has been shown that this divergence is due to sum rules on the Fourier transform of Eq.~\eqref{eq:Hr} and its derivatives not being satisfied exactly as $q\rightarrow 0$.\cite{TormeyThesis}  Related observations have been made about RISM.\cite{Cummings1982} Modifying the form of the solvation potential to remove this divergence in TM theory has been done.\cite{Yagi2020} An algorithm that enforces the sum rules without changing the form of the solvation potential has been developed, though.\cite{TormeyThesis} In that way, if the potential has the HNC-like form of Eq.~\eqref{W-HNC}, say, this algorithm allows it to reduce to its proper form\cite{DCM94} in the limit that the true potentials are weak. Further research into this question and improving the solvation potential itself are welcome.

\begin{acknowledgments}
We thank David Heine for helpful discussions.
\end{acknowledgments}

\section*{Author declarations}
The authors have no conflicts to disclose.

\appendix
\section{\label{app} Simulation algorithms}
As mentioned in Sec.~\ref{sec:numerical} above, one- and two-molecule simulations were employed to solve TM theory for the PE liquid. Details are given here.

\subsection{\label{sec:onechainmc}One-molecule importance sampling}
A one-molecule MC simulation using importance sampling was used to compute the intramolecular pair correlation function $\Omega(r)$. The algorithm is discussed only briefly here as a detailed explanation has been given elsewhere.\cite{PCG01} First, an estimate for the solvation potential, $w(r)$, was obtained either as an initial guess or as an output in the computation step for $g(r)$. See Figure \ref{figTCSolutionFlow} above. An initial chain conformation $\Re$ was then generated, it being consistent with the polyetheylene chain model of Sec.~\ref{sec:polychainmodel} above. 

Then a new chain conformation was generated using the standard pivot algorithm.\cite{MS88} Each pivot step consists of one attempted bending move and one attempted torsional move. For the bending move, a bond angle was picked randomly and one of its arms was rotated by a random angle
within $\pm20^\circ$ around the axis perpendicular to the angle
plane. For the torsional move, a dihedral angle is picked randomly
and the shorter arm is rotated by a random angle within
$\pm180^\circ$ around the axis created by the two center sites of
the dihedral. The chain energy of this new conformation was then compared with that of the former one with the winner decided by the Metropolis algorithm.  In this manner a sequence of conformations were generated. The auto-correlation functions of the bending and torsion energy of a single $N_m=96$ chain indicate a correlation time of $400$ pivot steps.\cite{HWCG03} Chain conformations were then saved
every $400th$ step. After a sufficiently large set was saved, usually $N_s= 5000$, then $\Omega(r)$ was obtained by averaging over all of them using Eq.~\eqref{omega-sol}. 

The time taken for a single one-molecule simulation is estimated to scale as $O(N_{sw}N_m^2)$, where $N_{sw}$ is the total number of MC sweeps (steps), and $N_m$ is the number of monomers per chain as defined in Sec.~\ref{sec:polychainmodel} above. Here, $N_{sw} = 400N_s$. 

\subsection{\label{sec::directsample}Two-molecule direct sampling}
The two-molecule equation, Eq.~\eqref{gr-sol}, has been evaluated previously by simulation for hard-sphere polymer chains by using a solvation potential derived from weighted density functional theory.\cite{YFS01} Direct sampling was used.

The use of direct sampling is suggested by the form of Eq.~\eqref{gr-sol} as an average of the Boltzmann factor $\exp({-{V}^{(2)}(\Re_1,\Re_2)})$ over the configurations of the two chains, while holding a site of type $k$ on chain 1 and a site of type $k'$ of chain 2 a fixed distance $r$ apart. A literal interpretation was implemented in Yethiraj et al.,\cite{YFS01} but was found to be expensive time-wise.  Here, an improved algorithm was used, which is able to collect data simultaneously on all pairs of sites between two chains, rather than on just the pair held fixed. In that way, the speed of computation is increased by a factor of $O(N_m^2)$. The algorithm is as follows.

First, a set of single chain conformations, $\{\Re\}$, is generated using the one-molecule MC simulation algorithm in Sec.~\ref{sec:onechainmc} above. Then the center of mass (c.o.m.) of each conformation is placed at the origin, and all are rotated randomly to achieve a more uniform sampling. From this set, conformations for chains 1 and 2 are randomly chosen. The c.o.m. of chain 2 is then displaced by a distance $z$ along the $z$-axis, $z$ chosen from the same grid of equally spaced points as the radial coordinate $r$. Denote the resultant chain configurations as $\Re_1$ and $\Re_2$, and their total state, $X = \{\Re_1,\Re_2\}$. For a given c.o.m. displacement $z$, this procedure is repeated $N_{tms}$ times to generate a set  ${\bf X}(z) = \{X\}$ of two-molecule configurations. 

Note that since the states are generated with the points $z$ being equally spaced, the total set, ${\bf X}_{tot}=\{{\bf X}(z)\}$, does not fill configuration space uniformly. To make the set quasi-uniform, add a factor of $z^2$ to the weight of each state. In that way, the total weight of a state $j$ is proportional to $(z^{(j)})^2\exp[-V^{(2)}(\Re_1^{(j)},\Re_2^{(j)})]$, where $z^{(j)}$ is the c.o.m. displacement associated with that state.

To compute $g(r)$, find all states in ${\bf X}_{tot}$ that have positions ${\bf r}_{11\alpha}^{(j)}$ and ${\bf r}_{21\beta}^{(j)}$ ($k=1$ denotes the $\rm CH_2$ site type here) separated by a distance $r$, add their weight to a running sum, and then when done normalize that sum by a term proportional to the volume of configuration space sampled. In that way,
\begin{equation}
\label{eq:grdirect}
g(r) = f(r)/n(r),
\end{equation}
where
\begin{eqnarray}
f(r)=&&\sum_j\sum_{\alpha,\beta}
(z^{(j)})^2 \exp\left [-V^{(2)}(\Re_1^{(j)},\Re_2^{(j)}) \right] \nonumber \\
&&\times\delta(r- \vert{\bf r}_{11\alpha}^{(j)} - {\bf r}_{21\beta}^{(j)}\vert),
\label{eq:frdirect}
\end{eqnarray} 
and the sum over $j$ runs over all $N_{tms}$ states for a given displacement $z$ and all $z$.
 The normalization function, $n(r)$, is the same as $f(r)$ but without the exponential factor. 

If the molecules were rigid, Eqs.~\eqref{eq:grdirect} and \eqref{eq:frdirect} would be a mapping from the molecular representation of the radial distribution function to the site-site one.\cite{Fries1994,Ishizuka2013}  Adding flexibility to the molecules, as done here, then merely increases the size of the configuration space being sampled.
 
Note that if $g(r)$ is determined on a grid of size $r_{max}$, then the maximum relative displacement of the molecule c.o.m., $z_{max}$, must be at least $r_{max} + d_{mol}$, where $d_{mol}$ is the maximum size of the liquid molecules. This ensures that all site-site pairs are sampled properly for relative distances near $r_{max}$. 

Note also that if $\Omega(r)$ is computed by reweighting a set of conformations from a previous one-molecule simulation, then this set of single chain conformations should also be reweighted in the computation for $g(r)$. However, since the one-molecule simulation was done more than once in computing $g(r)$, and the criterion for reweighting was conservative, the error in not doing so was determined to be negligible and so was ignored. 

Direct sampling appears to compute the long wavelength properties with less noise than the short wavelength. In that way, to obtain less noisy behavior for $g(r)$ for small $r$, $N_{tms}$ was sometimes set larger, to 20000, say, but the simulation result for the smaller $N_{tms}$ was always used for the initial guess. A statistical analysis of the errors associated with this and other sampling methods is discussed in Li.\cite{HuiminThesis}

With $g(r)$ and thus $h_{tm}(r)$ determined by Eq.~\eqref{eq:grdirect}, a new estimate for $c(r)$ and thus $w(r)$ is obtained using the methods of Sec.~\ref{sec:numerical} above.  It was important here that the random number generator be reinitialized with the same seed for each iteration involving a two-molecule simulation so as to minimize random noise from affecting the convergence of $g(r)$. In that manner, convergence could be attained for {\it any} reasonable value of $N_{tms}$, so other means were used to ensure that it was large enough to give good statistics.  

The time taken for a single two-molecule simulation is estimated to scale as $O(N_{tms}N_m^2N_r)$. Here, $N_{tms}$ was chosen to be equal to the number, $N_s$, of saved conformations in the one-molecule simulation. So with the number of $r$-grid points $N_r$ used here, the ratio of times for the two-molecule relative to the one-molecule simulation, $\tau_{2c}/\tau_{1c}\sim 2048/400 \sim 5$. Multi-threading can be implemented easily within direct sampling, and here ten threads were used. With this number and the one-molecule simulation running single threaded, then $\tau_{2c}/\tau_{1c}\sim 1/2$. In practice it was found that this ratio was accurate. The advantage of the two-molecule simulation compared with a full MD atomistic simulation can be seen here as the time for the former scales only linearly with system size $r_{max} = N_r\Delta r$, while the latter scales at best as $O(r_{max}^3)$.

%


\bibliography{huiminRef}


%
\end{document}